\documentclass[aps,prl,twocolumn,floatfix,amsmath,amssymb]{revtex4}
\usepackage[varg]{txfonts}
\usepackage[final]{graphicx}
\usepackage{amsmath,amssymb,bbold,bm}

\begin{document}

\newcommand{\bk}{{\bf k}}
\newcommand{\bp}{{\bf p}}
\newcommand{\bv}{{\bf v}}
\newcommand{\bq}{{\bf q}}
\newcommand{\tbq}{\tilde{\bf q}}
\newcommand{\tq}{\tilde{q}}
\newcommand{\bQ}{{\bf Q}}
\newcommand{\br}{{\bf r}}
\newcommand{\bR}{{\bf R}}
\newcommand{\bB}{{\bf B}}
\newcommand{\bA}{{\bf A}}
\newcommand{\bK}{{\bf K}}
\newcommand{\cS}{{\cal S}}
\newcommand{\vd}{{v_\Delta}}
\newcommand{\tr}{{\rm Tr}}
\newcommand{\rslash}{\not\!r}
\newcommand{\bs}{{\bar\sigma}}

\title{Giant proximity effect in a phase-fluctuating superconductor}

\author{Dominic Marchand, Lucian Covaci, Mona Berciu and Marcel Franz}
\affiliation{Department of Physics and Astronomy,
University of British Columbia, Vancouver, BC, Canada V6T 1Z1}
\date{\today}

\begin{abstract}
When a tunneling barrier between two superconductors is formed by a normal 
material that would be a superconductor in the absence of phase fluctuations, 
the resulting Josephson effect can undergo an enormous enhancement. We establish
this novel proximity effect by a general argument as well as a numerical 
simulation and argue that it may underlie recent experimental observations 
of the giant proximity
effect between two cuprate superconductors separated by a barrier made of the
same material rendered normal by severe underdoping.  
\end{abstract}
\maketitle

Josephson effect \cite{josephson1} -- the ability of Cooper pairs to coherently 
tunnel between two nearby superconductors --  represents one of the most
spectacular manifestations of the electron pairing paradigm that underlies
the BCS theory of superconductivity. If the barrier between the superconductors
is formed by an insulating material (or a vacuum) the tunneling current is 
controlled by the overlap between the Cooper pair wave functions that extend
into the empty space between the
superconductors. When the barrier is made
out of a normal metal (SNS tunneling) then the supercurrent can be much 
enhanced due to the proximity effect \cite{prox}.
In essence, local superconducting order is induced inside the barrier which 
significantly enhances the distance over which pairs can tunnel.

The proximity effect is well understood and documented in conventional 
superconductors \cite{conv1,conv2}. In high-$T_c$ cuprates there now exists 
compelling 
experimental evidence for anomalously large proximity effect when the barrier
is formed by an underdoped cuprate that would be in its normal state if studied
in isolation \cite{kabasawa1,tarutani1,kasai1,bozovic2,decca1,bozovic1}. The critical 
currents of such junctions have been reported
to exceed the expectations based on conventional theories by more than two
orders of magnitude. While the early results raised some suspicion of being 
contaminated by various extrinsic effects, the most recent data \cite{bozovic1} 
on very high quality films under closely controlled conditions leave little doubt that the effect
is intrinsic and that it represents a qualitatively new type of Josephson 
tunneling. Previous theoretical attempts to elucidate this effect were mostly 
based on modeling inhomogeneous barriers using conventional mean-field methods
\cite{kresin,dagotto,covaci} but they could not account 
for purely intrinsic effects.

In this Letter we formulate a theory of a new type of tunneling between two 
superconductors that occurs when the barrier
is formed by an {\em unconventional normal metal}. The latter is characterized 
as a superconductor that has lost its phase rigidity due to phase fluctuations.
According to one school of thought \cite{uemura1,emery1,randeria1,fm1,balents1,ft1} 
it is precisely this type of an unconventional normal metal that appears in the pseudogap state of 
cuprate superconductors \cite{timusk1} used as a barrier in the above 
experiments \cite{bozovic2,decca1,bozovic1}. Recent experimental insights into the pseudogap 
phase \cite{pasler1,corson1,ong1,ong2,lem1} 
seem to confirm the existence of vortices well above the critical temperature, 
supporting the phase fluctuation paradigm and calling for a description of
the Josephson tunneling processes in these exotic phases.

In the following we demonstrate, by a general argument and by extensive 
numerical simulations, that  the Josephson 
tunneling in this situation (which we hereafter refer to as `SPS' 
tunneling) is greatly enhanced compared to the SNS. We show that in SPS the 
dependence of the junction critical current on both the temperature and the 
barrier thickness is {\em qualitatively different} from the SNS
case. The most striking difference is that in one particular regime we find 
{\em logarithmic} dependence of the junction critical temperature $T_{\rm eff}$ 
on the junction width $d$. At $T<T_{\rm eff}$ this  
allows the pairs to tunnel over vastly longer distances in accord with 
experiment.

The standard model describing a phase-fluctuating superconductor 
is defined by the XY Hamiltonian
\begin{equation}
H_{\rm XY}=
-{1\over 2}\sum_{\langle ij\rangle}J_{ij}\cos({\varphi}_i-{\varphi}_j).
\label{hxy}
\end{equation}
Here $\varphi_i$ represents the phase of the superconducting order parameter 
on the site $\br_i$ of a $D$-dimensional square lattice and $J_{ij}$ are 
Josephson couplings between the neighboring sites $\br_i$ and $\br_j$.
Classical \cite{stroud1,carlson1} as well as the quantum 
\cite{paramekanti1,herbut1,iyengar1,mel1} 
versions of this model have been employed previously to study phase 
fluctuations in cuprates. Although quantum fluctuations may be
important
in cuprates, to demonstrate the effect in the simplest possible setting,
we focus here on the effect of classical thermal fluctuations.

\begin{figure}[t]
\includegraphics[width = 8.4cm]{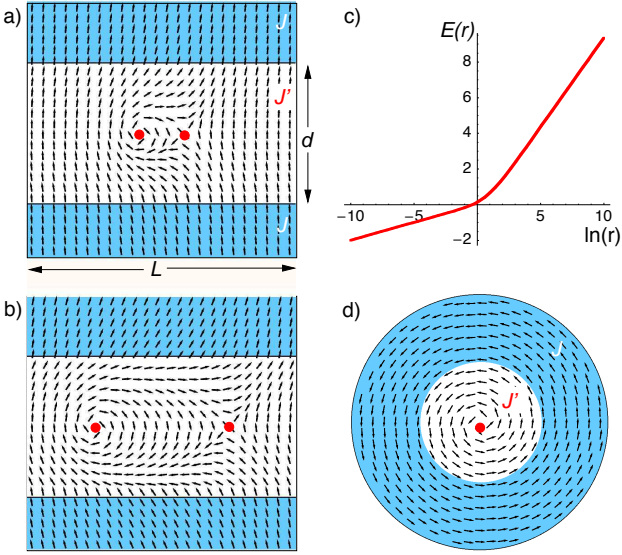}
\caption{(Color online) Superconducting phase distribution in the $J$-$J'$-$J$ 
junction
with $J'/J=0.3$ and a vortex-antivortex pair with size a) $r=0.3d$ and
b) $r=d$.  
c) Energy of the vortex-antivortex pair, Eq.\ (\ref{en}), in units of $J$ as a 
function of the pair size in units of $d$, for $J'/J=0.2$. d) Disk
geometry used to estimate the 
 vortex unbinding temperature.
}
\label{fig1}
\end{figure}

In the spatially uniform situation, $J_{ij}=J$, it is well known that the XY
model undergoes a superconductor to normal transition at the critical 
temperature
\begin{equation}
T_c=cJ,
\label{tc}
\end{equation}
where we took $k_B=1$. In 2 dimensions the transition 
is of the Kosterlitz-Thouless (KT) type \cite{kt1}, driven by the unbinding of 
vortex-antivortex pairs. The standard KT argument applied to the continuum XY
model gives $c=\pi/2$ while numerical simulations of the lattice model 
(\ref{hxy}) yield $c\simeq 0.93$ \cite{teitel1}. In order to model the 
proximity effect we consider the above XY Hamiltonian in the $J$-$J'$-$J$ 
geometry 
illustrated in Fig.~\ref{fig1}(a): two superconductors characterized by $J_{ij}=J$
separated by a strip of width $d$ and $J_{ij}=J'<J$. This configuration, 
at temperatures $T_c'<T<T_c$, behaves as a Josephson junction similar to the one
studied in Ref.\ \cite{bozovic1} since, according to Eq.\ (\ref{tc}), the 
strip should be in the normal state in this regime.

In this configuration the proximity of bulk superconductors will prevent 
vortex-antivortex pairs in the strip from unbinding at $T_c'$, leading
to anomalously large proximity effect. In order to see this consider a 
single pair of size $r \ll d$ inside the strip. The phase gradient is largely 
confined to within the  strip, Fig. \ref{fig1}(a), and thus the energy of such 
pair will be  $E_{va}(r)\simeq J'\ln(r)$. Once $r$ exceeds $d$, however, the 
phase gradient necessarily spills into the regions outside the strip, 
Fig. \ref{fig1}(b), and the pair becomes energetically
more costly. In the limit $r \gg d$ we expect $E_{va}(r)$ to approach $J\ln(r)$
and indeed this is borne out in a more detailed calculation summarized below.
Since the KT transition occurs when $r \to \infty$, i.e.\ vortices
become free, we may expect that the transition temperature in this geometry
will be controlled by $J$ and not by $J'$. The critical temperature
and the critical current of such an SPS junction will thus be significantly
enhanced.

Results presented in Fig.\ref{fig1} are based on the well known mapping \cite{map} of
the continuum version of Hamiltonian (\ref{hxy}), 
\begin{equation}
H_{\rm XY}=
{1\over 2}\int \textrm{d}\br^2 J(\br)(\nabla\varphi)^2,
\label{hxy_c}
\end{equation}
onto a 2D problem in electrostatics of point charges
(representing vortices) in the dielectric medium characterized by a dielectric
constant $\epsilon(\br)\sim J(\br)^{-1}$. The phase configurations (related to
the electric field vector) and the energy of the vortex-antivortex configuration
can be obtained by the method of image charges. For vortex-antivortex pair lying
on the symmetry axis of the strip the energy acquires a simple form,
\begin{equation}
E_{va}(r)/J'=\ln r + \sum_{j=1}^{\infty}\alpha^j\ln(1+r^2/j^2),  
\label{en}
\end{equation}
with $\alpha=(J-J')/(J+J')$ and $r$ measured in units of $d$. It is easy to verify that 
Eq.\ (\ref{en}) indeed implies the asymptotic behavior stated above and shown in
Fig.~\ref{fig1}(c).

We now proceed to estimate the vortex unbinding temperature $T_{\rm eff}$ 
\cite{kt1} in a 
system consisting of two superconductors characterized by $J$ and $J'$. For 
simplicity we consider the disk geometry sketched in Fig.~\ref{fig1}(d). The key 
advantage of this geometry is that, using the continuum Hamiltonian  
(\ref{hxy_c}), we can calculate the energy of a vortex placed at the center 
of the disk exactly. By symmetry it is easy to see that $|\nabla\varphi|=1/r$
and it follows that the vortex energy is
\begin{equation}
E_v=\pi\left[J'\ln(d/\xi) + J\ln(L/d)\right],
\label{ev}
\end{equation}
where $\xi$ is the vortex core cutoff and $d$, $L$ are inner and outer radii
respectively.
We shall henceforth assume that (\ref{ev}) remains approximately valid even when
the vortex is placed slightly off-center. The entropy of the 
single-vortex configuration can be estimated as $S_v\simeq\ln(L/\xi)^2$ and 
the vortex unbinding temperature is then obtained by examining the free energy 
of the system $F=E_v-TS_v$. This yields
\begin{equation}
T_{\rm eff}\simeq{\pi\over 2}J\left[1-\left(1-{J'\over J}\right){\ln (d/\xi)\over \ln (L/\xi)}
\right].
\label{tc1}
\end{equation}
The above formula 
is physically reasonable: it interpolates smoothly between the limiting cases
$d\to \xi$ and $d\to L$, giving $T_c=(\pi/2)J$ and $(\pi/2)J'$ respectively, in 
accord with Eq.\ (\ref{tc}). On the other hand we do not expect Eq.\ (\ref{tc1})
to remain accurate for $J'\ll J$. Indeed $J'\to 0$ constitutes a
singular limit: here we expect $T_{\rm eff}=T_c$ if $d=0$,
whereas $T_{\rm eff}=0$ for all finite $d$. This discontinuous behavior is unlike the linear
decrease of $T_{\rm eff}$ from $T_c$ to $T_c'=0$ predicted by
Eq. (\ref{tc1}) when $J'\to 0$.

The key implication of the above estimate is 
that the critical temperature should scale with the ratio of {\em logarithms}
of $d$ and $L$. Such an unusual scaling is a direct consequence of the non-local
nature of the phase field generated by a vortex and is much more general than
the crude treatment presented above may suggest. Specifically, we shall 
establish below by detailed numerical simulations that the logarithmic scaling 
(\ref{tc1}) also applies to the strip geometry of Fig.~\ref{fig1}(a), and thus 
by extension, to the experimental setup of Refs.\
\cite{bozovic2,decca1,bozovic1}. Since the critical current
of a junction also scales with its $T_c$ this establishes the advertised 
anomalous behavior of the SPS junction.

In order to validate the above considerations we now investigate 
the proximity effect systematically using numerical simulation.
We employ a version of the Monte Carlo method in which we
first map the XY Hamiltonian (\ref{hxy}) onto a bond-current model 
\cite{wallin} using a high-temperature expansion and then deploy the `worm algorithm' \cite{prokofev}, with only
minor complications due to the inhomogeneity of $J_{ij}$. This method is
well suited for our needs as it allows for 
efficient evaluation of the main quantity of interest, the helicity modulus 
$\Upsilon$ and is known to circumvent problems due
to critical slowing down near the transition. $\Upsilon$   
measures the response of the system to an externally imposed phase 
twist and its relevance follows from the fact that
it is proportional to the critical current $j_c$ the system can 
sustain before going normal \cite{map}. 

To check the validity of our algorithm, we first studied the homogeneous
system with $d=0$ in 2D. Our results in this limit are in excellent agreement
with those of Ref.\ \cite{teitel1} and, as $L$ gets large, exhibit a clear 
approach towards the  universal jump in $\Upsilon(T)$ expected near the KT 
transition.
\begin{figure}[t]
\includegraphics[height = \columnwidth,angle=-90]{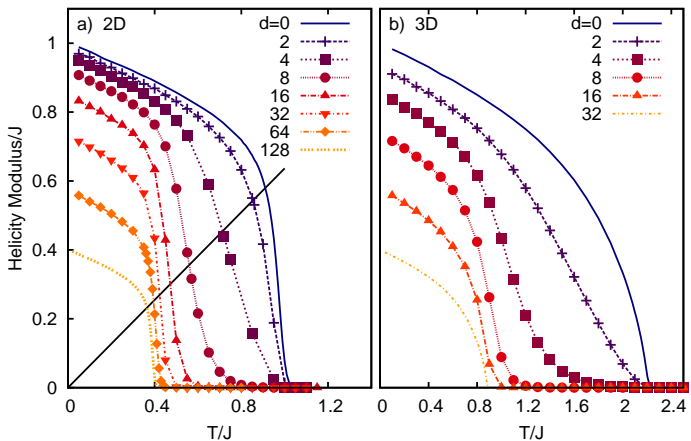}
\caption{(Color online) Helicity modulus as a function of temperature 
of the (a) 2D system with $L=128$ and (b) 3D system with
$L=32$ and different sizes of the barrier $d$ with $J'/J=0.4$.}
\label{fig2}
\end{figure}

We consider next the 2D strip geometry illustrated in Fig.~\ref{fig1}(a), 
for which we fix
$J'/J=0.4$ and $L=128$ while varying the size of the junction $d$ (here and 
hereafter we measure  $d$ and $L$ in units of $\xi$). The helicity modulus in 
the direction perpendicular to the junction is calculated as a function of
temperature for a wide range of $d$. This is shown in Fig.~\ref{fig2}(a). We
observe a smooth evolution of $\Upsilon(T)$ between the two
homogeneous geometries ($d=0$ to $d=L$). An interesting aspect of this 
data is the behavior of  $\Upsilon(T)$ at $d\ll L$: the helicity modulus 
(and thus $j_c$) of the junction remains large at temperatures far exceeding 
$T_c'$. This is quite striking when one recalls that in this geometry all the 
supercurrent must pass through the region of small coupling $J'$ which would 
be in the normal state at $T>T_c'$ if studied in isolation. We may thus 
conclude that in the experimentally relevant regime $d\ll L$ the junction
critical current is controlled largely by the properties of the leads, as
expected on the basis of heuristic arguments presented above.

Similar behavior occurs in 3D as illustrated in Fig.~\ref{fig2}(b). We note 
that in 3D 
there is no universal jump at $T_c$; instead our simulation recovers the 
expected continuous behavior characterized by the 3D-XY exponent 
$\nu\simeq0.667$. 
We note that both 2D and 3D results in
Fig.~\ref{fig2} exhibit the characteristic {\em linear} $T$-dependence in the 
vicinity of the junction critical temperature that is ubiquitous in the 
experimental data \cite{bozovic1}.

\begin{figure}[t]
\includegraphics[height = \columnwidth,angle=-90]{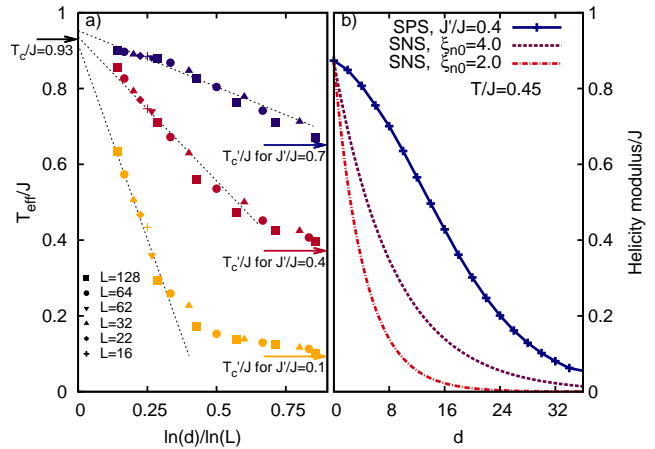}
\caption{(Color online) a) Junction critical temperature $T_{\rm eff}$ as a
function of $x=\ln{d}/\ln{L}$ for three different ratios  $J'/J=0.7,
0.4$ and 0.1 in 2D. Finite
size effects prevent these points from falling exactly on a smooth
curve. The arrows indicate the expected $T_c'$ values, reached when $d
\rightarrow L$. Both $d$ and $L$ are measured in units of $\xi$. b) 
Comparison between SPS model and conventional SNS proximity effect.}
\label{fig3}
\end{figure}

In order to quantify the proximity effect we consider the junction 
critical temperature $T_{\rm eff}$ defined in 2D by the intersection of 
$\Upsilon(T)$ with the line with slope equal to $2/\pi$ \cite{teitel1}.
In Fig.~\ref{fig3}(a) the
logarithmic dependence of $T_{\rm eff}$ on $x=\ln{d}/\ln{L}$
expected from Eq.~(\ref{tc1})  is seen to hold for small $x$. The slope
increases with decreasing $J'/J$, in accord with Eq.~(\ref{tc1}), although it
is quantitatively somewhat larger than predicted, presumably due to the 
differences between the strip geometry and the cylindrical geometry used to 
derive Eq.~(\ref{tc1}). For $x\rightarrow 1$,  $T_{\rm eff}$
tends to $T_c'$, as expected. In the limit $J'/J\rightarrow 0$ we see
a pronounced departure from the linear variation between  $T_c$ and $T_c'$
predicted by Eq. (\ref{tc1}). This is consistent with our expectation
that Eq. (\ref{tc1}) fails to describe this singular limit. Moreover, our
numerical results demonstrate   how the discontinuous jump of $T_{\rm
  eff}$ from $T_c$ to $T_c'=0$,
expected for $J'=0$, is approached continuously as $J' \rightarrow 0$.

We interpret the above numerical data as being in qualitative agreement with
our heuristic picture of the giant proximity effect through a phase fluctuating
superconductor in 2D. Notably, the agreement is excellent in the experimentally 
relevant regime of the narrow barrier $x\ll 1$. 

In real superconductors the logarithmic 
interaction between vortices, which gives rise to the above phenomena, is cut
off exponentially at length scales exceeding the magnetic penetration depth
$\lambda$, or the effective `Pearl' length $\lambda_{\rm eff}=\lambda^2/h$ in a 
2D film of thickness $h$. When modeling a real superconductor one should thus 
replace $L$ in all formulas by  $\Lambda=\max(\lambda,\lambda_{\rm eff})$ 
except for very small junctions ($L<\Lambda$) in which case $T_{\rm eff}$
will depend on the size of the macroscopic leads $L$ as indicated
by Eq.\ (\ref{tc1}). In cuprates  $\kappa\equiv\lambda/\xi\simeq 10^2-10^4$
 and the barrier thickness $d$ is typically of the order of 10-200\AA. Thus, the
junctions of Ref.\ \cite{bozovic1} are in the limit of relatively small 
$x=\ln (d/\xi)/\ln\kappa$ and our considerations should apply.

In 3D point-like vortices are replaced by vortex loops. These lead to
similar non-local phase gradients as vortex-antivortex pairs in 2D and 
it is thus to be expected
that the enhancement of the proximity effect will persist in 3D  SPS junctions.
This is indeed confirmed by Fig.~\ref{fig2}(b). Within the error bars our 
3D numerical data hint at logarithmic dependence  of $T_{\rm eff}$ on $d$ 
similar to that in Fig~\ref{fig3}(a) but we do not currently 
have a simple heuristic picture for this dependence. Detailed account of 
our analysis will be given elsewhere \cite{marchand1}.

Ref.\ \cite{bozovic2} describes a Bi-2212/Bi-2201/Bi-2212
junction 123\AA~ thick, with $T_{\rm eff}\simeq 50$K, over 3 times higher than 
$T_c'\simeq 15$K of the Bi-2201 film. In
 the standard theory of SNS tunneling \cite{prox,conv1,conv2} the critical 
current 
\begin{equation}\label{jc}
j_c\sim e^{-d/\xi_n}.
\end{equation}
In the 2D clean limit $\xi_n=\xi_{n0}\sqrt{T_c'/(T-T_c')}$ 
with $\xi_{n0}= {1\over 2}\xi(\Delta_0/k_BT_c)$ and $\xi$ is the BCS coherence 
length of the order of tens of \AA~ in cuprates. One thus expects essentially
no supercurrent to 
flow through the above junction at temperatures significantly above $T_c'$ 
according to the
conventional theory. In the SPS scenario advocated in this Letter such 
enhancement is easily attainable due to the weak logarithmic dependence 
of $T_{\rm eff}$ on the barrier thickness $d$. Physically, this key difference 
stems from our assumption, rooted in extensive experimental evidence 
\cite{uemura1,pasler1,corson1,ong1,ong2,lem1}, that underdoped cuprates above $T_c$ behave
as phase-disordered superconductors. To further exemplify this contrast we 
compare in Fig.~\ref{fig3}(b) the $j_c$ dependence on the junction width   
obtained from our model with the conventional SNS theory Eq.\ (\ref{jc}).
We observe that for reasonable values of the BCS ratio 
($\Delta_0/k_BT_c\approx 4-8$ in cuprates) the SPS model implies vastly larger 
critical current than the conventional SNS theory. To the extent that
our predictions can be systematically verified, experimental observation
of the giant proximity effect can be viewed as a smoking gun evidence
for the phase fluctuation paradigm.

Finally, it is worth pointing out that our results could also be
relevant to suitably fabricated Josephson junction arrays as well as 
junctions of thin ferromagnetic films with easy-plane anisotropy 
and different exchange integrals (and thus,
different Curie temperatures). These are also described by the 
$XY$ model and our results apply unchanged, except that in ferromagnet 
the helicity modulus is replaced by the spin stiffness. Experimental 
measurements of
its dependence on the  thickness $d$ of the inside layer provide
another way to verify our predictions.

The authors are indebted to B. Seradjeh, S. Teitel, 
Z. Te\v{s}anovi\'c, N. Prokof'ev, B. Svistunov and E. Burovski for stimulating discussions and correspondence. This work 
was supported by NSERC, CIFAR, FQRNT and the A.P. Sloan Foundation.


\begin{thebibliography}{99}
\bibitem{josephson1} B.D. Josephson, Rev. Mod. Phys. {\bf 46}, 251 (1974).

\bibitem{prox} P.G. de Gennes,  Rev. Mod. Phys. {\bf 36}, 225 (1964).

\bibitem{conv1} G. Deutscher and P. G. De Gennes, in Superconductivity, ed.\ 
by R.D. Parks (Marcel Dekker, New York, 1969).

\bibitem{conv2} K. K. Likharev, Dynamics of Josephson Junctions and Circuits (Gordon and Breach, New York, 1986).

\bibitem{kabasawa1}U. Kabasawa {\em et al.}, Jpn. J. Appl. Phys. {\bf 30}, 1670 (1991). 

\bibitem{tarutani1} Y. Tarutani {\em et al.}, Appl. Phys. Lett. {\bf 58}, 2707 (1991).

\bibitem{kasai1} M. Kasai {\em et al.}, J. Appl. Phys. {\bf 72}, 5344 (1992).

\bibitem{bozovic2} I. Bozovic {\em et al.}, J.\ Supercond.\ {\bf 7}, 187 (1994).

\bibitem{decca1} R.S. Decca {\em et al.}, \prl {\bf 85}, 3708 (2000).

\bibitem{bozovic1} I. Bozovic {\em et al.}, \prl {\bf 93}, 157002 (2004).

\bibitem{kresin} V. Kresin {\em et al.}, Appl. Phys. Lett. {\bf 83} 722 (2003).

\bibitem{dagotto} G. Alvarez {\em et al.}, \prb {\bf 71}, 014514 (2005).

\bibitem{covaci} L. Covaci and F. Marsiglio, \prb {\bf 73}, 014503 (2006).


\bibitem{uemura1} Y.J. Uemura {\em et al.}, Phys. Rev. Lett {\bf 62}, 2317 (1989). 

\bibitem{emery1} V.J. Emery and S.A. Kivelson, Nature {\bf 374}, 434 (1995).

\bibitem{randeria1} M. Randeria, {\em Varenna Lectures} (cond-mat/9710223).

\bibitem{fm1} M. Franz and A.J. Millis, \prb {\bf 58}, 14572 (1998).

\bibitem{balents1} L. Balents, M.P.A. Fisher and C. Nayak, \prb {\bf 60}, 1654 (1999).

\bibitem{ft1} M. Franz and Z. Te\v{s}anovi\'c, \prl {\bf 87}, 257003 (2001).

\bibitem{timusk1} T. Timusk and B.W. Statt,  Rep. Prog. Phys. {\bf 62}, 61 (1999).


\bibitem{pasler1} V. Pasler {\em et al.}, \prl {\bf 81}, 1094 (1998).

\bibitem{corson1} J. Corson {\em et al.}, Nature {\bf 398}, 221 (1999).

\bibitem{ong1} Z.A. Xu {\em et al.}, Nature {\bf 406}, 486 (2000).

\bibitem{ong2} Y. Wang {\em et al.}, \prl {\bf 95}, 247002 (2005).

\bibitem{lem1} I. Hetel, T.R Lemberger and M. Randeria,  Nature Phys.  {\bf 3}, 
700 (2007).

\bibitem{stroud1} E. Roddick and D. Stroud, \prl {\bf 74}, 1430 (1995).

\bibitem{carlson1} E.W. Carlson {\em et al.}, \prl {\bf 83}, 612 (1999).

\bibitem{paramekanti1} A. Paramekanti, {\em et al.}, \prb {\bf 62}, 6786 (2000).

\bibitem{herbut1} I.F. Herbut and M.J. Case, \prb {\bf 70}, 094516 (2004).

\bibitem{iyengar1} M. Franz and A.P. Iyengar, Phys. Rev. Lett. {\bf 96}, 047007 (2006). 

\bibitem{mel1} A. Melikyan and Z. Tesanovic, \prb {\bf 71}, 214511 (2005).

\bibitem{kt1} J.M. Kosterlitz and D.J. Thouless, J. Phys. C {\bf 6}, 1181
(1973). 

\bibitem{teitel1} S. Teitel and C. Jayaprakash, \prb {\bf 27}, 598 (1983).

\bibitem{map} P. Minnhagen, Rev. Mod. Phys. {\bf 59}, 1001 (1987).

\bibitem{wallin} M. Wallin {\em et al.}, \prb {\bf 49}, 12115 (1994).

\bibitem{prokofev} N. V. Prokof'ev and B. V. Svistunov,  \prl {\bf 87}, 160601 (2001).

\bibitem{marchand1} D. Marchand {\em et al}. (unpublished).


\end{thebibliography}
\end{document}